\documentclass[aps,prd,twocolumn,a4paper,superscriptaddress,nofootinbib]{revtex4-1}
\usepackage{graphicx,caption,subcaption,amsmath,amsfonts,amssymb,multirow,extarrows,bm,acronym,float}
\usepackage[colorlinks,linkcolor=blue,citecolor=blue,urlcolor=blue ]{hyperref}
\floatstyle{plaintop}
\restylefloat{table}
\newcommand{\fs}{\mathcal{F}\text{-statistic}}

\newcommand{\Mpc}{\,{\rm Mpc}}
\newcommand{\Msun}{\,{\rm M}_\odot}

\newcommand{\PKU}{Kavli Institute for Astronomy and Astrophysics, Peking University, Beijing 100871, People's Republic of China}
\newcommand{\DLUT}{School of Physics, Dalian University of Technology, Liaoning 116024, People's Republic of China}
\newcommand{\PKUA}{Department of Astronomy, School of Physics, Peking University, Beijing 100871, People's Republic of China}
\newcommand{\NAOCAS}{National Astronomical Observatories, Chinese Academy of Sciences, Beijing 100012, People's Republic of China}
\def\mnras{MNRAS}            % Monthly Notices of the RAS
\def\prd{Phys.~Rev.~D}       % Physical Review D
\def\prl{Phys.~Rev.~Lett}    % Physical Review Letters
\begin{document}

\title{Reanalyzing the ringdown signal of GW150914 using the $\mathcal{F}$-statistic method}

\author{Hai-Tian Wang}
\email{wanght9@dlut.edu.cn}
\affiliation{\DLUT}
\author{Ziming Wang}
\affiliation{\PKUA}
\affiliation{\PKU}
\author{Yiming Dong}
\affiliation{\PKUA}
\affiliation{\PKU}
\author{Garvin Yim}
\affiliation{\PKU}
\author{Lijing Shao}
%% \email{lshao@pku.edu.cn}
\affiliation{\PKU}
\affiliation{\NAOCAS}

\date{\today}

\begin{abstract}
The ringdown phase of a gravitational wave (GW) signal from a binary black
hole merger provides valuable insights into the properties of the final
black hole and serves as a critical test of general relativity in the
strong-field regime. A key aspect of this investigation is to determine
whether the first overtone mode exists in real GW data, as its presence
would offer significant implications for our understanding of general
relativity under extreme conditions.
To address this, we conducted a reanalysis of the ringdown signal from GW150914, using the newly proposed $\mathcal{F}$-statistic method to search for the first overtone mode. Our results are consistent with those obtained through classical time-domain Bayesian inference, indicating that there is no evidence of the first overtone mode in the ringdown signal of GW150914.
However, our results show the potentiality of utilizing the $\fs$ methodology to unearth nuanced features within GW signals, thereby contributing novel insights into black hole properties.
%% Leveraging the strengths inherent to the $\mathcal{F}$-statistic, our findings corroborate evidence for the presence of the first overtone mode in GW150914, with a Bayes factor of $50$ under the assumption that the ringdown signal initiates from peak amplitude. 
%% The evidence is further supported by performing an injection test.
%% The discovery and substantiation of the first overtone mode carries significant implications for understanding black hole physics and in facilitating gravitational wave detections. 
%% We perform a test of no-hair theorem and find no deviation from general relatively.
\end{abstract}

\maketitle
%%%%%%%%%%%%%%%%%%%%%%%%%%%%%%%%%%%%%%%%%%%%%%%%%%%%%%%%%%%%%%%%

\acrodef{GW}{gravitational wave}
\acrodef{LIGO}{Laser Interferometer Gravitational-Wave Observatory}
\acrodef{LVC}{LIGO-Virgo Collaboration}
\acrodef{NR}{numerical relativity}
\acrodef{TD}{time-domain}
\acrodef{BH}{black hole}
\acrodef{BBH}{binary black hole}
\acrodef{GR}{general relativity}
\acrodef{PN}{post-Newtonian}
\acrodef{SNR}{signal-to-noise ratio}
\acrodef{PSD}{power spectral density}
\acrodef{PDF}{probability density function}
\acrodef{ACF}{auto-covariance function}
\acrodef{IMR}{inspiral-merger-ringdown}
\acrodef{QNMs}{quasinormal modes}

\section{Introduction}\label{sec:intro}

%% ringdown signal and GW150914
Throughout the initial three observing runs conducted by the LIGO-Virgo-KAGRA Collaboration, in excess of $90$ events have been identified. Of these, the vast majority come from the inspiral of binary black holes (BBHs) \citep{LIGO_PRX2019, LIGO_O3a_PRX2020, KAGRA:2021vkt}. 
Following a BBH's violent collision, the resulting remnant \ac{BH} oscillates and emits \acp{GW} until it reaches equilibrium as per the no-hair theorem \citep{Hawking:1971vc, PhysRevLett.34.905}. 
The corresponding \ac{GW} signal emitted during this phase is termed the ``ringdown'' signal and can be mathematically represented as a superposition of \ac{QNMs} \citep{Schw_PRD_Vishveshwara1970, GW_APJL_Press1971, QNM_APJ_Teukolsky1973}. 
These QNMs can be further decomposed into spin-weighted spheroidal harmonics with angular indices $(\ell,m)$, each comprised of a series of overtones denoted by $n$ \citep{Berti:2009kk}. 
The fundamental mode $(2,2,0)$ pertains to the case where $\ell=|m|=2$ and $n=0$, which constitutes the dominant mode during the ringdown. This mode is particularly significant as it lasts longer than higher overtone modes ($n\geq 1$) and possesses a significantly larger amplitude compared to higher multipoles ($\ell > 2$ and $\ell \geq |m|$) for comparable-mass binaries.

%% GW150914 and overtone modes
GW150914, identified as the inaugural BBH event \citep{gw150914_PRL2016}, exhibits a postpeak \ac{SNR} reaching $14$ \citep{2019PhRvL.123k1102I,2021PhRvD.103l2002A,2021arXiv211206861T,Test_GR_150914}, thereby rendering it suitable for ringdown analysis. 
The absence of higher multipoles in GW150914's ringdown signal has been supported by previous investigation~\cite{PRD_Carullo2019,Gennari:2023gmx}. 
Nevertheless, certain studies such as those testing the no-hair theorem \citep{2019PhRvL.123k1102I, 2021PhRvD.103b4041B} necessitate at least two modes within the ringdown signal. 
Furthermore, \citet{Hirano:2024fgp} found that including higher overtone modes improves constraints on gravity theories.
Consequently, investigating potential evidence of overtone modes in GW150914's ringdown signal is deemed crucial. 
In a study on overtones, \citet{Overtone_PRX_Giesler2019} successfully fitted \ac{NR} ringdown signals with a waveform that included $7$ overtone modes and discovered these modes to be well-fitted even when assuming that the start of the ringdown signal occurs from the peak amplitude of the \ac{NR} waveform. 
However, signals surrounding this peak should theoretically belong to a highly nonlinear region rather than a perturbative one where QNM calculations are applicable; thus implying that fitting within this region could potentially be unphysical.
For instance, there are also some studies \citep{Baibhav:2023clw,Nee:2023osy,Zhu:2023mzv,Clarke:2024lwi} which argue that higher overtones $(n>2)$ overfit the transient radiation and nonlinearities close to the merger.
Recently, \citet{Giesler:2024hcr} confirmed the existence of higher overtone and second-order modes using highly accurate numerical waveforms. They further argue that, despite being mathematically classified as nonlinear, second-order modes can be computed by applying the linear perturbation theory.
This calls for further investigation into the role of overtone modes within real \ac{GW} data analysis for \ac{BH} spectroscopy.

%% the first overtone in GW150914
In the analysis of GW150914's ringdown signal from its peak at a sampling rate of $2048$ Hz, \citet{2021PhRvL.127a1103I} finds evidence for the first overtone mode. Conversely, when analyzing the same ringdown signal but with a higher sampling rate of $16384$ Hz, \citet{2022PhRvL.129k1102C} concludes that noise dominated any evidence for this overtone mode. 
For further details, one can refer to the follow-up comment by \citet{Isi:2023nif} and the corresponding reply by \citet{Carullo:2023gtf}.
The discrepancy among these findings can be attributed to their respective methods of estimating noise in the data, the importance of which has been well studied in Refs.~\citep{Wang:2023mst,Siegel:2024jqd}.
Upon implementing a more accurate method for noise estimation, consistent results were obtained across different sampling rates by \citet{Wang:2023mst}, who confirmed no discernible evidence supporting the existence of this first overtone mode.
Results in Ref.~\citep{Wang:2023mst} have been validated by a subsequent study \citep{Correia:2023bfn}.
%% However, it is important to note that these conclusions are constrained by limitations inherent to the methodology employed in Ref.~\citep{Wang:2023mst}. 
%% To address this, the approach used here introduces only two additional parameters to model the first overtone mode.
%% Specifically, this approach introduces two additional parameters related to modeling this first overtone mode.

%% from mode cleaning to Fs
To date, various methods have been developed for performing ringdown analysis, including the time-domain method \citep{Isi:2021iql, Isi:2022mhy, Wang:2023mst}, the frequency-domain method \citep{Finch:2021qph, Finch:2022ynt, CalderonBustillo:2020rmh}, rational filters \citep{Ma:2023cwe, Ma:2023vvr}, $\fs$ \citep{Wang:2024jlz}, and machine learning approaches \citep{Crisostomi:2023tle, Srinivasan:2024uax}.
In Refs.~\citep{Ma:2023cwe, Ma:2023vvr}, evidence was discovered for the first overtone mode in the GW150914 ringdown signal. 
This discovery was made using a distinct method known as rational filters and yielded a Bayes factor reaching $600$. 
A significant advantage of this approach is that it does not expand parameter space when incorporating additional modes into the ringdown waveform. 
To validate the result from \citet{Ma:2023cwe}, we have devised another novel methodology for ringdown analysis. 
This new method is based on the concept of $\fs$ \citep{Wang:2024jlz}, which was originally formulated for continuous \ac{GW} signals \citep{Jaranowski:1998qm,Cutler:2005hc,Dreissigacker:2018afk,Sieniawska:2019qnx} and subsequently applied to extreme mass-ratio inspiral signals \citep{Wang:2012xh}. %% Now, it has been utilized in the analysis of ringdown signals \cite{Wang:2024jlz}.
In brief, in $\fs$ parameters like amplitudes and phases of \ac{QNMs} are all analytically maximized so the inclusion of more modes does not lead to an expansion in parameter space. 
Moreover, this method proves more robust without loss in information, for instance, information about the inclination.

The structure of this paper is as follows. The concept of $\fs$ is introduced in Sec.~\ref{sec:methods}. Our primary findings, derived from the analysis of the GW150914 ringdown signal and injection test utilizing the $\fs$, are presented in Sec.~\ref{sec:bayes}. Finally, a concise summary and discussion are provided in Sec.~\ref{sec:con}. Unless stated otherwise, we employ geometric units where $G=c=1$ throughout this paper.

%%%%%%%%%%%%%%%%%%%%%%%%%%%%%%%%%%%%%%%%%%%%%%%%%%%%%%%%%%%%%%%%%
\section{The $\mathcal{F}$-statistic}\label{sec:methods}
%--------------------------------------------------------

In accordance with \ac{GR}, we postulate that the eventual remnant of
GW150914 is a Kerr \ac{BH}. Different from the conventional \ac{TD} method
employed in Ref.~\citep{Wang:2023mst}, each mode of the ringdown waveform
is restructured as follows:
\begin{equation}
\begin{aligned}
B^{\ell mn,1}=&A_{\ell mn}\cos\phi_{\ell mn},\\
B^{\ell mn,2}=&A_{\ell mn}\sin\phi_{\ell mn},\\
g_{\ell mn,1}=&\left[F^+\cos(2\pi f_{\ell mn}t)+F^{\times}\sin(2\pi f_{\ell mn}t)\right]\\
&\times {}_{-2}Y_{\ell m}(\iota, \delta)\exp \Big(-\frac{t}{\tau_{\ell mn}} \Big),\\
g_{\ell mn,2}=&\left[-F^{+}\sin(2\pi f_{\ell mn}t)+F^{\times}\cos(2\pi f_{\ell mn}t)\right]\\
&\times {}_{-2}Y_{\ell m}(\iota,\delta)\exp \Big(-\frac{t}{\tau_{\ell mn}} \Big),
\label{eq:rin_split}
\end{aligned}
\end{equation}
where $F^{+,\times}$ denotes the antenna pattern functions that depend on both sky location and the \ac{GW} polarization angle.
The damping frequency ($f_{\ell mn}$) and damping time ($\tau_{\ell mn}$)
are determined by two factors: the final mass\footnote{In this work, we do
not involve the transformation between the source frame and the detector
reference frame, and all quantities are defined in the detector frame.
Therefore, the final mass means the redshifted final mass.}
 ($M_f$) and final spin
($\chi_f$) of the remnant. The inclination angle is denoted as $\iota$,
while $\delta$ signifies the azimuthal angle and is fixed to zero in our
calculation. 
 We define $B^{\mu}$ and $g_{\mu}$ as
unified notations, where $B^{\mu}$ represents the set of components $\big\{
B^{\ell mn,1}, B^{\ell mn,2} \big\}$ and $g_{\mu}$ represents the corresponding
functions $\big\{ g_{\ell mn,1}, g_{\ell mn,2} \big\}$, indexed by the
multi-indices $(\ell, m, n)$. Consequently, the recorded ringdown in a
single detector can be expressed as $h(t) = B^\mu g_\mu(t)$. 

In accordance with Refs.~\citep{PRD_Carullo2019,2022PhRvL.129k1102C}, this study does not incorporate contributions from higher harmonics, such as the $\ell=|m|=3$ and $n=0$ mode, which is potentially present in sources exhibiting larger mass ratios \citep{2020PhRvD.102h4052L,Capano:2021etf,Capano:2022zqm}. Higher-order overtone modes ($n \geq 2$) are also excluded from consideration. The approximation of using spherical harmonics in place of spheroidal harmonics is employed in Eq.~\eqref{eq:rin_split}, a method whose efficacy has been empirically validated in Ref.~\citep{Overtone_PRX_Giesler2019}. 

\ac{GW} signals are intrinsically intertwined with noise, characterized by its autocovariance matrix $\mathcal{C}$ in the \ac{TD}. The determination of this autocovariance matrix is contingent upon its \ac{ACF}. Utilizing the same methodology as employed in Ref.~\citep{Wang:2023mst}, we estimate the ACF from GW data. We define the inner product of two distinct signals ${h}_1$ and ${h_2}$ as follows: 
\begin{equation}\label{eq:inner1}
\langle {h}_1\vert {h}_2\rangle=\vec{h}_1^{\intercal}\mathcal{C}^{-1}\vec{h}_2,
\end{equation}
where the arrow notation $\vec{h}$ indicates that it is a discrete
time series sampled from the continuous signal $h(t)$. 

In the \ac{TD}, the log-likelihood can be expressed as
\begin{equation}
\ln \mathcal{L}=-\frac{1}{2}\langle d - h | d - h \rangle +C_0\,,
\label{eq:ll}
\end{equation}
where $C_0$ is a constant and independent of $d$ and $h$. %% , omitted in the discussion hereafter.
Following the process shown in \citet{Wang:2024jlz} and substituting $h =
B^\mu g_\mu$, the log-likelihood can be rewritten as
\begin{equation}
    \ln \mathcal{L} = {\cal F} - \frac{1}{2} \Big[\big(B^{\mu}- \hat{B}^{\mu}\big) M_{\mu\nu} \big(B^{\nu}- \hat{B}^{\nu}\big)  + \langle d | d \rangle \Big] \,,
    \label{eq:ll2}
\end{equation}
where $M_{\mu\nu} \! = \langle g_\mu |
g_\nu \rangle$ and
 $\hat{B}^{\mu}\! = (M^{-1})^{\mu\nu} s_{\nu}$ with $s_{\nu} \! = \langle d | g_\nu \rangle$. Note
 that these three quantities do not depend on $B^{\mu}$. The first term of Eq.~\eqref{eq:ll2} is called the $\fs$,
\begin{equation}
    {\cal F}= \frac{1}{2} s_{\mu} (M^{-1})^{\mu\nu} s_{\nu}+C_0,\label{eq:fs}
\end{equation}
which only depends on the data $d$ and the signal of each QNM mode $g_\mu$.
In the Bayesian inference, we use the likelihood defined by Eq.~\eqref{eq:fs}.
Furthermore, a detailed study based on Eq.~\eqref{eq:ll2} can be found in \citet{Dong:2025igh}.

When conducting parameter estimation, $g_\mu$, like the $\fs$, depends
on other source parameters ${\bm \theta}$ rather than on the QNM amplitudes and
phases ${\bm B} = \big\{B^1,\cdots,B^{2N}\big\}$, where $N$ is the number of
considered QNM modes. In GW ringdown analysis, ${\bm \theta}$ embodies seven
parameters, namely $\big\{\text{RA}, \text{DEC}, t_c,\psi,\iota,M_f,\chi_f\big\}$, which represent two sky position angles, geocentric reference time,
polarization angle, inclination angle, final mass and final spin,
respectively. It is also necessary to fix $(\text{RA}, \text{DEC},
t_\mathrm{c}, \psi)$ based on other analyses, such as results derived from
a comprehensive \ac{IMR} analysis. Therefore, in this work we take ${\bm
\theta} = \big\{M_f,\chi_f,\iota\big\}$. In
addition, for a network consisting of $N_{\rm det}$ detectors, 
$s_{\mu}$ and $M_{\mu\nu}$ should be replaced by the summation of
$s^1_{\mu}+s^2_{\mu}+\cdots+s^{N_{\rm det}}_{\mu}$ and
$M^1_{\mu\nu}+M^2_{\mu\nu}+\cdots+M^{N_{\rm det}}_{\mu\nu}$, respectively~\citep{Wang:2012xh}.

In contrast to the case of continuous GWs \citep{Prix:2009tq}, Eq.~\eqref{eq:ll2} does not reduce to Eq.~\eqref{eq:fs} when flat priors are assumed for ${\bm B}$, due to the nonconstancy of the determinant of $M_{\mu\nu}$. In Ref.~\citep{Prix:2016tv4},\footnote{We thank the anonymous referee for pointing out this reference, which we noticed after their kind reminder.} the author performs ringdown analyses using a formulation similar to Eq.~\eqref{eq:ll2}. In the present study, we adopt $\fs$, as defined in Eq.~\eqref{eq:fs}, for the ringdown analysis. However, direct computation of the Bayes factor is not feasible, as we do not assume any priors on ${\bm B}$. Therefore, we introduce a new criterion, the $\fs$ information criterion (FIC), defined as, 
\begin{equation} 
\mathbf{FIC}=\log\int\exp\left(\mathcal{F}-\frac{1}{2}\langle d|d\rangle\right)\cdot p(\bm\theta)d\bm\theta-K,\label{eq:fic} 
\end{equation} 
where $K$ represents the dimension of ${\bm B}$, i.e., $K=2N$ in this case. The definition of the FIC is both straightforward and meaningful: it reduces to the logarithmic Bayes factor when $K=0$ and to half the negative Akaike information criterion (AIC) \citep{Akaike1973InformationTA,Burnham2003ModelSA} when the dimension of $\bm \theta$ is zero. We find that it performs well in the current study, as detailed in Sec.~\ref{sec:bayes}. Future work will further investigate its application in other contexts, such as in the full inspiral-merger-ringdown analysis. To compare the performance of two models with $N=1$ and $N=2$, we use the difference $\mathbf{FIC}^{N=2}_{N=1} \equiv \mathbf{FIC}_{N=2} - \mathbf{FIC}_{N=1}$.

%% For the model comparison, we can calculate the Bayes factor between two
%% models consisting of different numbers of QNM modes. In Eq.~\eqref{eq:ll2}, the rewritten log-likelihood using $\fs$ has a
%% Gaussian form for $\bm B$, which brings a significant advantage in the
%% evidence calculation. Especially, choosing the flat prior $\pi({\bm B}) =
%% \pi_B$, one can analytically marginalize over $\bm B$ as
%% \begin{align}
%%     \mathcal{Z}&=\!\int\! \pi({\bm \theta}) \pi({\bm B}) e^{\mathcal{F}({\bm \theta})-\frac{1}{2}\langle d |d\rangle-\frac{1}{2}(B^{\mu}\!-\!\hat{B}^{\mu})M_{\mu\nu}(B^{\nu}\!-\!\hat{B}^{\nu})}\mathrm{d}{\bm \theta}\mathrm{d}{\bm B}\notag \\
%%     &=\pi_B(2\pi)^N e^{-\frac{1}{2}\langle d |d\rangle}\int\pi({\bm \theta})\sqrt{{\rm det}\big(M^{-1}\big) }e^{\mathcal{F}({\bm \theta})}\mathrm{d}{\bm \theta}.
%% \end{align}
%% Hence, one only needs to calculate a 3-dimensional integral for the
%% evidence. The Bayes factor $\mathcal{B}$ of two different models $N=1,2$
%% can be computed from their corresponding evidences,
%% $\mathcal{B}^{N=2}_{N=1}=\mathcal{Z}_{N=2}/\mathcal{Z}_{N=1}$.

%--------------------------------------------------------
\section{Results of Bayesian inferences}\label{sec:bayes}
%--------------------------------------------------------

%--------------------------------------------------------
\begin{figure*}
\centering
\begin{subfigure}[b]{0.48\linewidth}
\centering
\includegraphics[width=\textwidth,height=8cm]{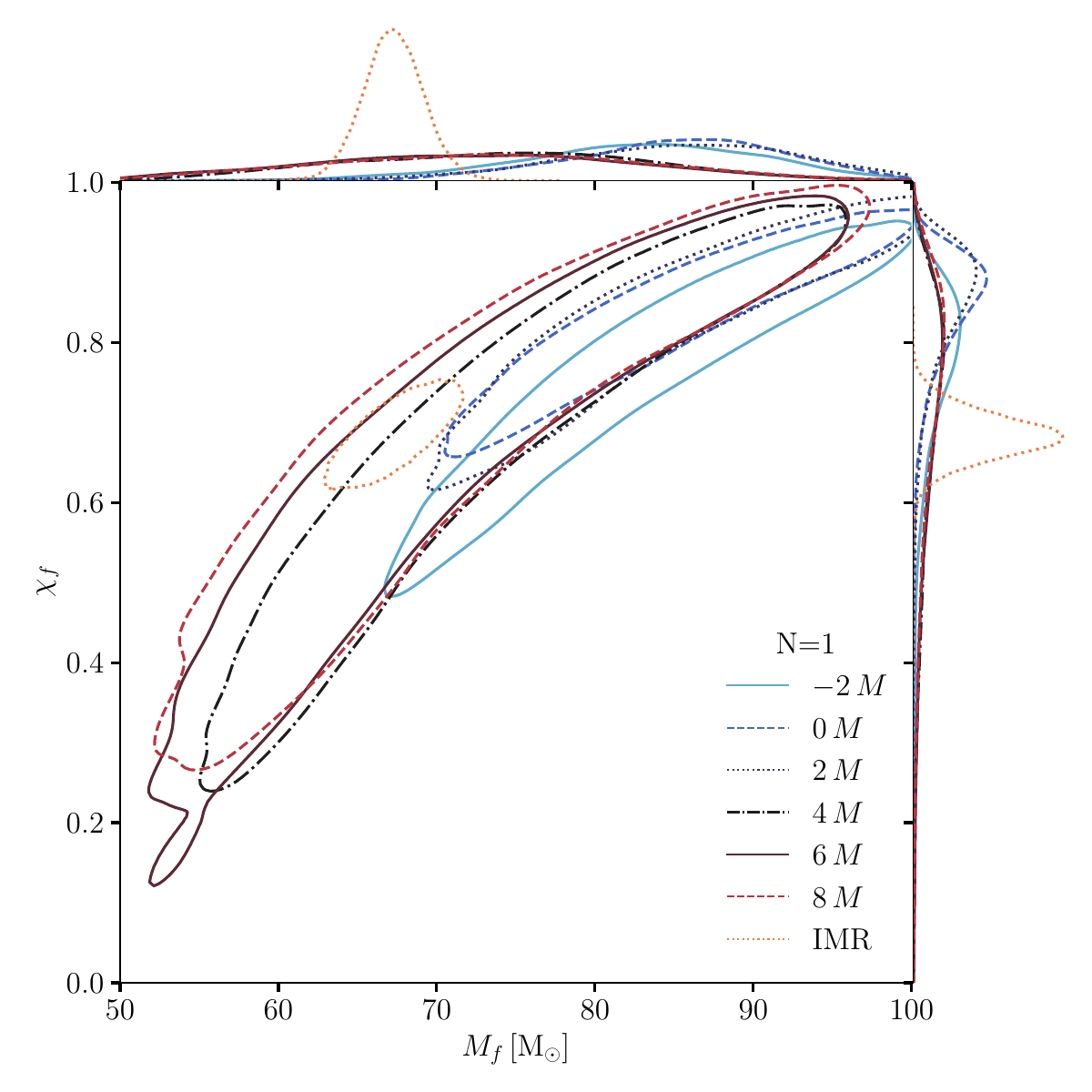}
\end{subfigure}%
\begin{subfigure}[b]{0.48\linewidth}
\centering
\includegraphics[width=\textwidth,height=8cm]{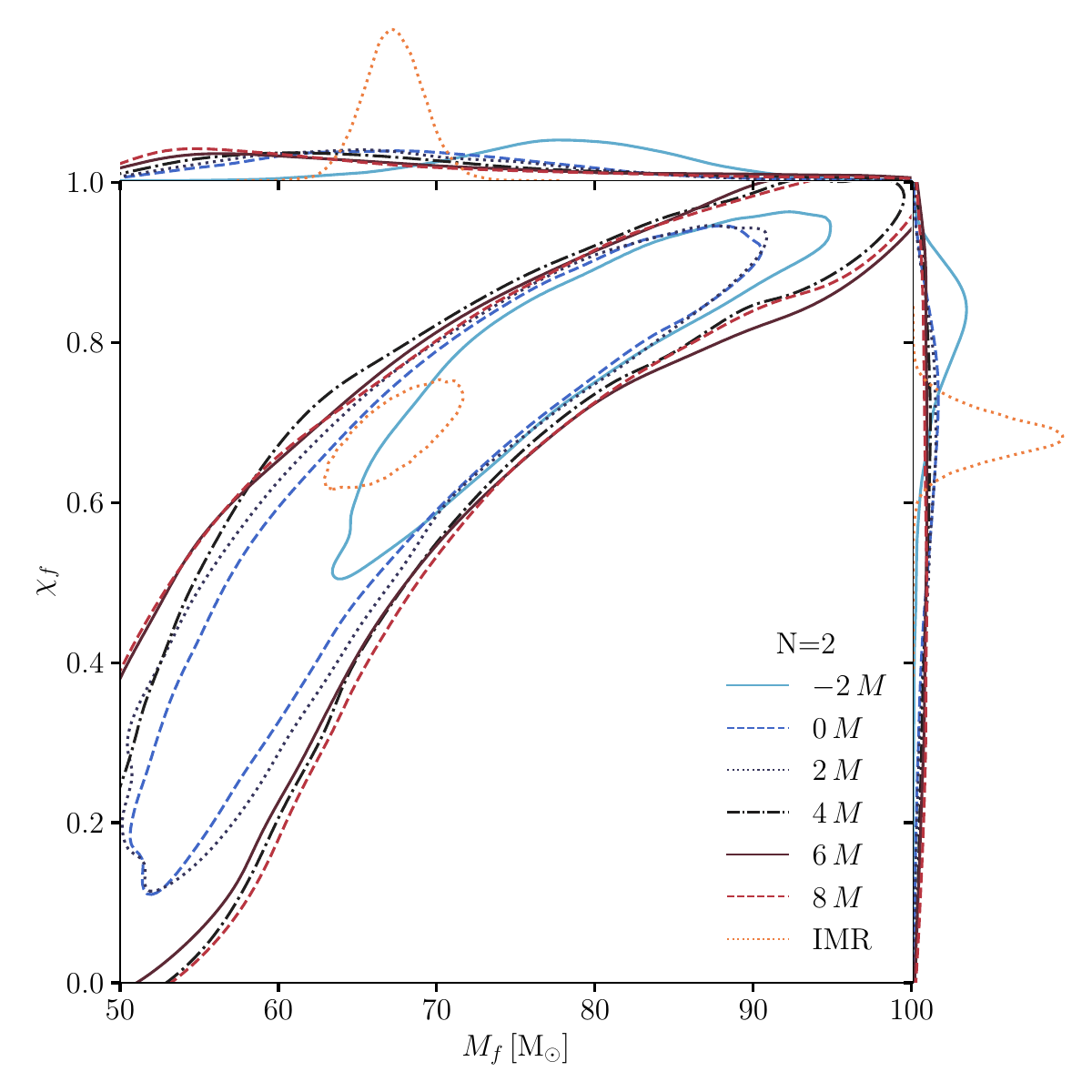}
\end{subfigure}%
\caption{
The posterior distributions of the final mass $M_f$, and the
final spin $\chi_f$, of the GW150914 remnant were obtained
utilizing the $\fs$ method. The contours depicted in different colors
correspond to different start times $\Delta t$, ranging from $-2\,M$ to
$8\,M$. Results derived from fundamental-mode-only analyses are displayed
on the left panel, while those incorporating the first overtone mode are
presented on the right panel. Contours are shown at the $90\%$ credible
level. The full \ac{IMR} analysis results
are indicated by dotted orange contours \citep{2021PhRvD.103l2002A}.
Additionally, the marginal posterior distributions for both
$M_f$ and $\chi_f$ are shown in their respective top and right
panels.
}\label{fig:joint_plots}
\end{figure*}
%--------------------------------------------------------

The Bayesian inference is conducted utilizing the {\sc Bilby} package
\citep[v2.1.1;][]{Ashton:2018jfp} and the {\sc Dynesty} sampler
\citep[v2.1.2;][]{Dynesty_MNRAS_Speagle2020}, incorporating $1000$ live
points and a maximum threshold of $1000$ Markov chain steps. In alignment
with Ref.~\citep{Wang:2023mst}, $\Delta t=t_c-t_{\rm ref}$ discretely spans from
$-2\,M$ to $8\,M$, where the constant $M$ uses the final remnant
mass of GW150914, $M=68.8\Msun$. 
Our selection of these start times is also informed by the methodologies used in other studies, such as \citep{2022PhRvL.129k1102C}, where similar considerations were made.
The geocentric time of GW150914 is denoted by $t_{\rm ref}=1126259462.40854$ GPS while $t_c$ signifies the
hypothesized commencement time of the ringdown signal. For the parameters
$M_f$ and $\chi_f$, we use the same priors as in Ref.~\citep{Wang:2023mst}.
The prior $\iota$ is set as $\pi(\iota)\propto \sin \iota$.
%% The application of $\fs$ requires flat priors for $\bm B$ \citep{Prix:2009tq}, and we set the maximum of the amplitude $A_{\ell mn}$ as $5\times 10^{-20}$.

In the pursuit of evidence for the first overtone mode, we execute Bayesian
inferences on two distinct models. The first model considers only the
fundamental mode within the ringdown waveform, denoted as $N=1$. Our second
model incorporates both the fundamental and first overtone modes, denoted
as $N=2$.  For each mode, we perform Bayesian inferences with different start times range from $-2M$ to $8M$.
A summary of posterior distributions for the final mass ($M_f$)
and final spin ($\chi_f$) can be found in Fig.~\ref{fig:joint_plots}.

In the $N=2$ model, a smaller $\Delta t$ imposes more stringent constraints
on the remnant. The joint distributions exhibit bias when $\Delta t=-2\,M$,
due to the inclusion of signals from nonlinear regions. Optimal constraints
are achieved at $\Delta t=0\,M$, where $M_f=67.7^{+16.5}_{-13.1} \, {\rm M}_\odot$ and
$\chi_f=0.66^{+0.21}_{-0.41}$ at a credible level of $90\%$. However, for
the $N=1$ model, joint distributions scarcely encompass median values from
\ac{IMR} until $\Delta t$ reaches $6\,M$. This is consistent with previous
studies indicating that overtone modes dominate during early stages of the
ringdown signal \citep{Overtone_PRX_Giesler2019}. At $\Delta t=8\,M$,
remnant constraints are given by $M_f=73.0^{+16.4}_{-17.0} \, {\rm M}_\odot$ and
$\chi_f=0.75^{+0.16}_{-0.42}$ with the $90\%$ credible level. 
As a comparison, using the traditional TD method, \citet{Wang:2023mst} reported $M_f=68.5^{+16.7}_{-13.9} \, {\rm M}_{\odot}$ and
$\chi_f=0.65^{+0.24}_{-0.46}$ at the $90\%$ credible level with a logarithmic Bayes factor of $-0.2$ for the $N=2$ model when $\Delta t=0M$.
We find that constraints derived from the use of the $\mathcal{F}$-statistic align closely with those obtained via traditional TD method. 
Similarly, the FIC result from the $\mathcal{F}$-statistic method is $-0.3$, which is in close agreement with the logarithmic Bayes factor of $-0.2$ reported by the traditional TD method.
%% However, this conclusion may suffer from different priors on amplitudes and phases. 
%% We will investigate the effects of different priors for the $\fs$ method in future work.
%% Despite this, we can anticipate the potential of the $\mathcal{F}$-statistic method to detect subtle features within GW signals.
%% Constraints derived from the use of the $\fs$ marginally agree with those obtained via
%% traditional \ac{TD} methods \citep{Wang:2023mst}.

%--------------------------------------------------------
\begin{figure*}
\centering
\includegraphics[width=0.66\textwidth,height=4.5cm]{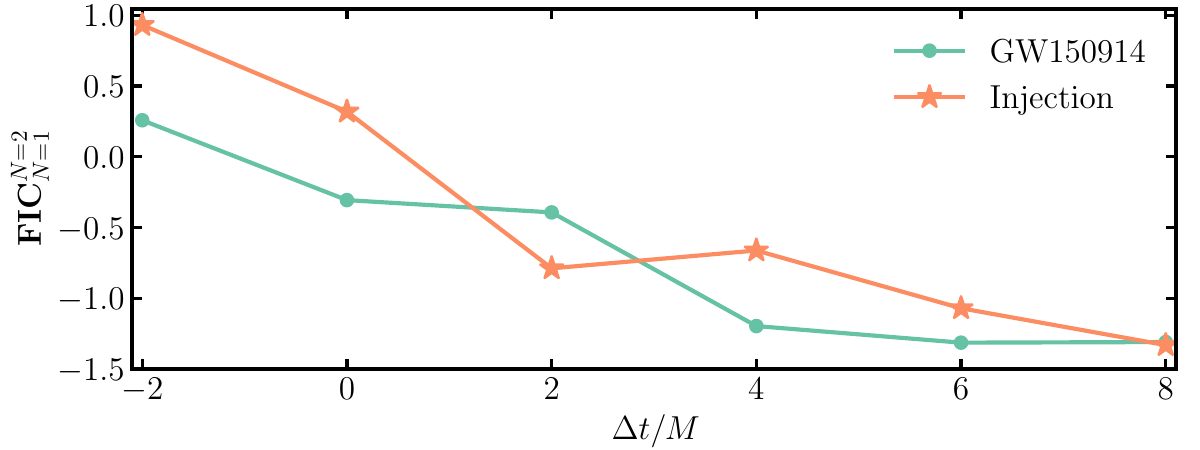}
\caption{
The FICs between models $N=2$ and $N=1$ for different
$\Delta t$. For comparative purposes, we also display results from injection tests (indicated by an orange line and ``$\star$'' markers), which are based on a GW150914-like NR waveform.
}\label{fig:bfs}
\end{figure*}
%--------------------------------------------------------

To ascertain the presence of the first overtone mode in the GW150914
signal, we show the FICs between models $N=2$ and
$N=1$ in Fig.~\ref{fig:bfs}. For $\Delta t=0\,M$, the FICs is
$\mathbf{FIC}^{N=2}_{N=1}=-0.3$, which does not support 
the model incorporating the first overtone mode. The FIC decreases
as $\Delta t$ increases because the analyzed strain data contain less of the ringdown signal as $\Delta t$ increases. 

To conduct a more in-depth analysis, we
run an injection test witn a \ac{NR} waveform SXS:BBH:0305,
which is similar to GW150914 and is found within
the Simulating eXtreme Spacetimes catalog \citep{2019CQGra..36s5006B}. The
waveform characterizes a source with a mass ratio of $0.82$
and a remnant with dimensionless spin $\chi_f=0.69$.
We fix the chirp mass to $31\Msun$ therefore fixing the constant $M$ in
this scenario to $M=68.2\Msun$. A luminosity distance of $490\Mpc$ is
utilized along with an inclination angle of $3\pi/4$, while other
parameters remain consistent with values established in the early study
\cite{Wang:2023mst}. The waveform corresponding to the spherical harmonics,
where both azimuthal and magnetic quantum numbers are equal ($l=|m|=2$), is
injected into Gaussian noise derived from GW data surrounding the GW150914
event. In this case, the postpeak signal's SNR is approximately $14.4$
which bears similarity to that of GW150914.

%--------------------------------------------------------
\begin{figure*}
\centering
\begin{subfigure}[b]{0.48\linewidth}
\centering
\includegraphics[width=\textwidth,height=8cm]{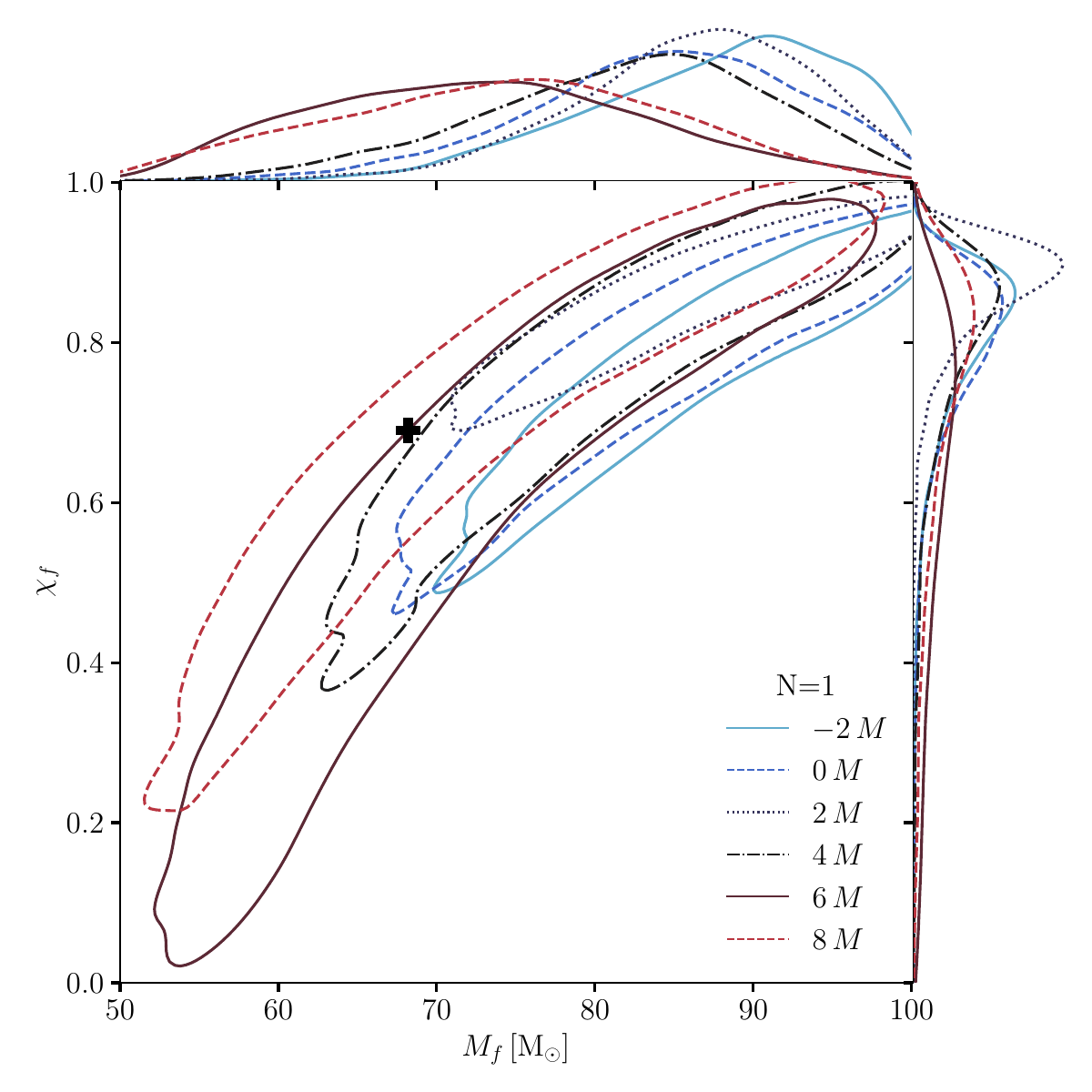}
\end{subfigure}%
\begin{subfigure}[b]{0.48\linewidth}
\centering
\includegraphics[width=\textwidth,height=8cm]{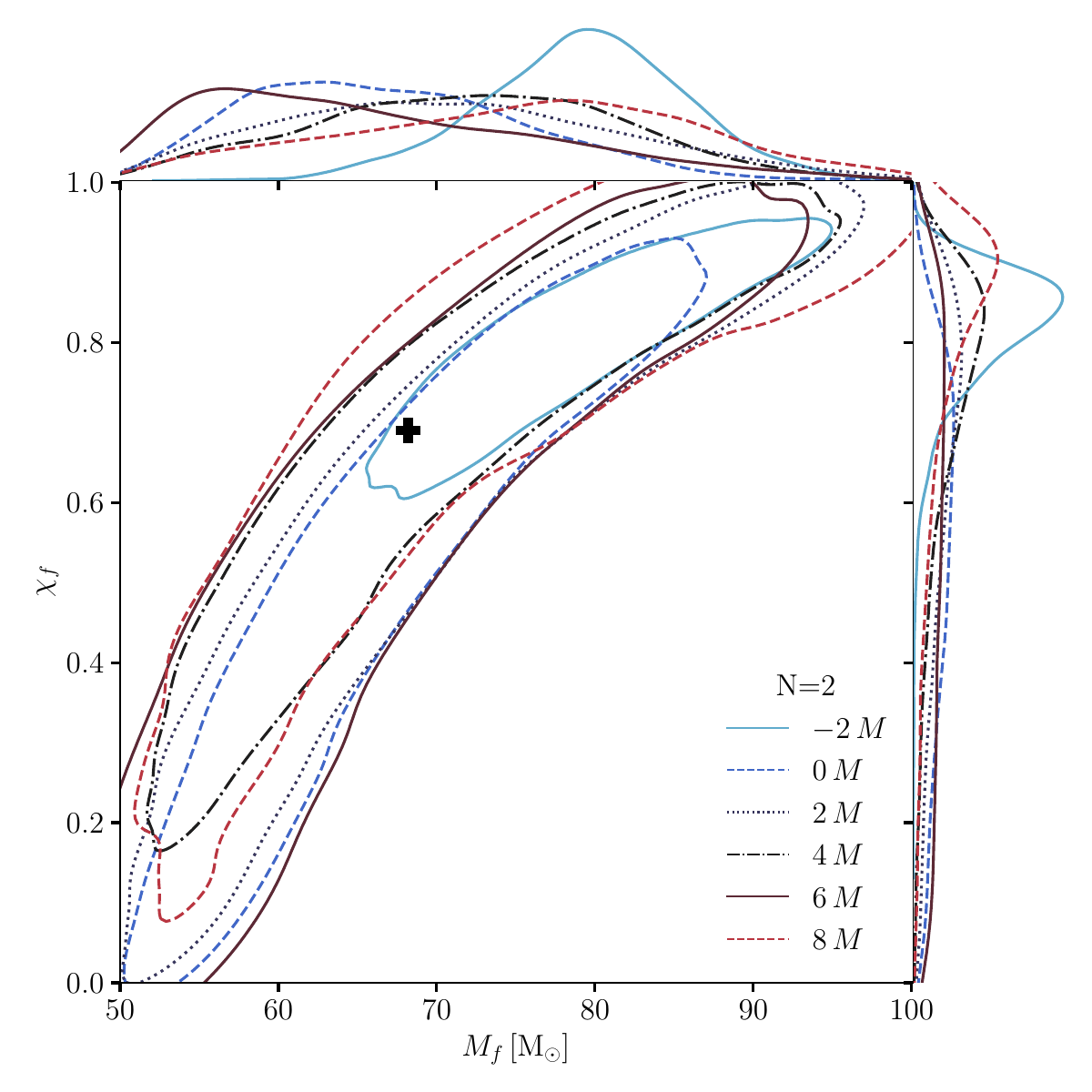}
\end{subfigure}%
\caption{Similar to Fig.~\ref{fig:joint_plots}, but from the injection test.
The black ``+" marker represents values of the final mass $M_f=68.2\Msun$ and final spin $\chi_f=0.69$ in the injection.
}\label{fig:inj}
\end{figure*}
%--------------------------------------------------------

Following the injection process, we execute TD Bayesian inferences
with the $\fs$ method for different $\Delta t$ values. The posterior
distributions for both redshifted final mass $M_f$ and final spin $\chi_f$
are shown in Fig.~\ref{fig:inj}. For models denoted as $N=1,2$, their joint
distributions bear resemblance to those depicted in
Fig.~\ref{fig:joint_plots}. The corresponding FICs are presented in Fig.~\ref{fig:bfs}.
When $\Delta t=0\,M$, the FIC is around $0.3$.
In the $N=2$ model, the final mass and final spin are $66.1^{+16.1}_{-11.5}\Msun$ and $0.55^{+0.29}_{-0.44}$
respectively with the $90\%$ credible level. 
In contrast, for the case where $\Delta t=8\,M$ with $N=1$, the constraints on both final mass ($M_f$) and final spin ($\chi_f$) are $74.3^{+16.2}_{-18.2}\Msun$ and $0.77^{+0.15}_{-0.46}$ respectively, also at a $90\%$ credible level.
This suggests that incorporating the first overtone mode into
the ringdown analysis could potentially enhance constraint precision slightly.
Overall, there is a substantial agreement between outcomes derived from injection tests when compared with those obtained from GW150914.

\section{Discussion and Conclusion}\label{sec:con}

We conducted a reanalysis of the ringdown signal from GW150914 utilizing the $\fs$ method, which possesses three primary advantages \citep{Wang:2024jlz}. Firstly, this approach ensures that the parameter space does not expand with an increased number of \ac{QNMs} incorporated into the ringdown waveform. Second, it operates more efficiently as ${\bm B}$ are analytically maximized over in the log-likelihood. Third, this method offers flexibility for extension to other research areas, including tests of the no-hair theorem. 
Leveraging these benefits allows us to effectively analyze GW150914's ringdown signal. 
%% Leveraging these benefits allows us to find evidence of the first overtone mode in GW150914's ringdown signal with a Bayes factor of $50$, under the assumption that said signal commences from its peak amplitude. Moreover, results of testing the no-hair theorem using GW150914's ringdown signal reveals no deviation from \ac{GR}.

We scrutinize the ringdown signal of GW150914 utilizing two distinct
models. The initial model encompasses solely the fundamental mode, denoted
as $N=1$. The second model incorporates both the first overtone mode and
the fundamental mode, represented as $N=2$. We execute Bayesian inferences
for each respective model with a total of six different start times range. 
For the case of $\Delta t=0\,M$ when $N=2$, the final mass and spin are
constrained to $M_f = 67.7^{+16.5}_{-13.2}\Msun$ and $\chi_f =
0.66^{+0.21}_{-0.41}$ respectively, at the $90\%$ credible level.
Comparing the case of $\Delta t=0\,M$ with that of $N=1$, it is observed
that the FIC is $\mathbf{FIC}^{N=2}_{N=1}=-0.3$
which does not support the presence of the first overtone mode in GW150914. 
To further substantiate this conclusion, an injection test was performed.
Adopting a GW150914-like NR waveform, the injection test yielded 
$\mathbf{FIC}^{N=2}_{N=1}=0.3$, consistent with the GW150914 case.

%% The primary source of discrepancy can be attributed to the method employed for noise estimation. It is anticipated that a higher degree of consistency would be achieved if the \ac{ACF} utilized in Refs.~\citep{Ma:2023cwe, Ma:2023vvr} were estimated following the same methodology as \citep{Wang:2023mst}. Nevertheless, this technical detail does not undermine our principal conclusion; namely, that an approach characterized by a smaller parameter space facilitates detection of the first overtone mode.
%% This research underscores not only the efficacy but also the potential of utilizing the $\fs$ methodology to unearth nuanced features within GW signals, thereby contributing novel insights into \ac{BH} properties.

%% This stems from the fact that in ringdown analyses, the start time -- typically inferred from \ac{IMR} analysis -- holds significant importance. 
%% Consequently, we anticipate that the $\fs$ can contribute towards a more accurate estimation of the start time.
The findings presented here show agreement with those in Refs.~\citep{2022PhRvL.129k1102C,Wang:2023mst,Correia:2023bfn}. 
Specifically, compared with constraints using the traditional TD method \citep{Wang:2023mst}, constraints provided by the $\mathcal{F}$-statistic method are slightly tighter. Therefore, we can anticipate the potential use of the $\mathcal{F}$-statistic method to detect subtle features within GW signals.

Undoubtedly, the analysis of the ringdown signal from additional \ac{GW} events utilizing the $\fs$ is crucial. 
The $\fs$ is more efficient since ${\bm B}$ can be analytically maximized.
Prior to this, our intention is to incorporate the $\fs$ into ringdown analyses for future detectors such as Einstein Telescope \citep{2010CQGra..27s4002P}, Cosmic Explorer \citep{2019BAAS...51g..35R}, Laser Interferometer Space Antenna \citep{LISA_arxiv2017}, TianQin \citep{TQ_2015,TianQin:2020hid}, Taiji \citep{1093nsrnwx116}, and DECIGO \citep{Kawamura:2020pcg}.
Furthermore, our proposed framework exhibits flexibility for extension to \ac{BH} spectroscopy based on \ac{NR} waveforms and future detector-identified events. It also serves as an effective instrument for testing \ac{GR} and constraining non-Kerr parameters (see e.g. Ref.~\cite{Gu:2023eaa}).
The data that support the figures of this article are openly available \citep{hai_tian_2024_14192256}.

\begin{acknowledgments}

We express our gratitude to the anonymous referee for their meticulous reading of the manuscript, as well as to Dicong Liang and Yi-Ming Hu for their insightful discussions.
This work was supported by the Beijing Natural Science Foundation (1242018), the National Natural Science Foundation of China (11991053), the National SKA Program of China (2020SKA0120300), the Max Planck Partner Group Program funded by the Max Planck Society.
H.-T Wang and L. Shao are supported by ``the Fundamental Research Funds for the Central Universities" respectively at Dalian University of Technology and Peking University. 

This research has made use of data or software obtained from the Gravitational Wave Open Science Center (gwosc.org), a service of LIGO Laboratory, the LIGO Scientific Collaboration, the Virgo Collaboration, and KAGRA~\cite{KAGRA:2023pio}. 
LIGO Laboratory and Advanced LIGO are funded by the United States National Science Foundation (NSF) as well as the Science and Technology Facilities Council (STFC) of the United Kingdom, the Max-Planck-Society (MPS), and the State of Niedersachsen/Germany for support of the construction of Advanced LIGO and construction and operation of the GEO600 detector. 
Additional support for Advanced LIGO was provided by the Australian Research Council. 
Virgo is funded, through the European Gravitational Observatory (EGO), by the French Centre National de Recherche Scientifique (CNRS), the Italian Istituto Nazionale di Fisica Nucleare (INFN) and the Dutch Nikhef, with contributions by institutions from Belgium, Germany, Greece, Hungary, Ireland, Japan, Monaco, Poland, Portugal, Spain.
KAGRA is supported by Ministry of Education, Culture, Sports, Science and Technology (MEXT), Japan Society for the Promotion of Science (JSPS) in Japan; National Research Foundation (NRF) and Ministry of Science and ICT (MSIT) in Korea; Academia Sinica (AS) and National Science and Technology Council (NSTC) in Taiwan of China.
\end{acknowledgments}

%%%%%%%%%%%%%%%%%%%%%%%%%%%%%%%%%%%%%%%%%%%%%%%%%%%%%%%%%%%%%%%%%

\bibliographystyle{apsrev4-1}
\bibliography{fs150914}

\end{document}